\documentclass[a4paper,11pt]{article}
\usepackage{pos}

\title{Baryon Transition Form Factors from Dynamical Coupled-Channel Analyses}

\author*[a]{Yu-Fei Wang}

\affiliation[a]{School of Nuclear Science and Technology, University of Chinese Academy of Sciences,\\
East Yanqi Road No.1, 101408 Beijing, China}

\emailAdd{wangyufei@ucas.ac.cn}

\abstract{In this talk, the electromagnetic transition form factors from the nucleon ground state to twelve $N^*$ and $\Delta$ states are exhibited and discussed. Those results are extracted through a comprehensive coupled-channel approach -- the Juelich-Bonn-Washington model, with the center-of-mass energy ranging from $1.13$ GeV to $1.8$ GeV, and the photon virtuality up to $8$ GeV$^2$. The extraction is based on $10^5$ electroproduction data points of $\pi N$, $\eta N$, and $K\Lambda$ channels, with additionally about $5\times 10^4$ data points in the hadronic sector as well as photoproductions as boundary conditions. The form factors are defined from the residues at the corresponding resonance poles in the multipole amplitudes. Uncertainties are also estimated by the exploration of the parameter space. The qualitative behavior of the transition form factors of $N(1440)$ and $\Delta(1232)$ here are in agreement with the previous studies, while for the other states, there has not been literature results that are also defined at the resonance poles. }

\FullConference{The 21st International Conference on Hadron Spectroscopy and Structure (HADRON2025)\\
27 - 31 March, 2025\\
Osaka University, Japan\\}

\begin{document}
\maketitle

Revealing the structures of hadrons is one of the fundamental tasks in physics. Especially, the $N^*$ and $\Delta$ states at intermediate energies encode rich information about the Quantum Chromodynamics (QCD), such as the confinement and chiral symmetry breaking~\cite{Mai:2022eur}. On the experimental side, photons are clean probes for the structures of those states. For example, photoproduction reactions of mesons off the nucleon~\cite{Ireland:2019uwn}, as a supplement to the $\pi N$ scatterings, help determining the states that do not strongly couple to $\pi N$. In addition, electroproduction reactions~\cite{Aznauryan:2012ba} provide an extra energy scale -- the photon virtuality $Q^2$. Specifically, electromagnetic transition form factors (TFFs) between the excited and the ground state baryons~\cite{Aznauryan:2011qj,Ramalho:2023hqd} are functions of $Q^2$ that indicate the
nature of the resonances like $N(1440)$~\cite{Burkert:2017djo}. 

However, the theoretical studies on the TFFs of the $N^*$ and $\Delta$ states are not easy due to the non-perturbative nature of QCD in the intermediate energy region. Efforts have been made in three aspects. First, the TFFs can be predicted at the quark level via quark models and Dyson-Schwinger approaches, see e.g. Refs.~\cite{Burkert:2017djo,Segovia:2015hra,Aznauryan:2018okk,Lu:2019bjs,Burkert:2025coj}. Lattice QCD calculations of the TFFs are also feasible~\cite{Agadjanov:2014kha,Lin:2020jmn}. Second, the TFFs can also be obtained through chiral effective theories and their extensions~\cite{Bernard:1996bi,Mai:2012wy,Ruic:2011wf,Hilt:2017iup}. Third, the TFFs can be extracted from experimental data with the help of phenomenological models. Unitary isobar models~\cite{Drechsel:2007if,Tiator:2011pw} are usually employed, with the real-valued TFFs depending on the Breit-Wigner parameters. Note that in principle, the TFFs should be independent of the model parametrization and the hadronic interaction channels, which is satisfied by a mathematically rigorous definition through the residues of the multipole amplitudes at the resonance poles. With secondary parametrizations and fits, one can access the TFFs at the pole in unitary isobar models~\cite{Tiator:2016btt}. Nevertheless, in dynamical coupled-channel (DCC) approaches~\cite{Doring:2025sgb}, the amplitudes are got by solving the dynamical scattering equations with analytic properties. Hence the resonances are automatically defined as the complex poles on the second Riemann sheet, with the complex-valued TFFs proportional to the residues. There have been only a few discussions on the TFFs from DCC models, for example the studies based on the ANL-Osaka model~\cite{Diaz:2006ios,Suzuki:2010yn,Kamano:2018sfb}. 

This talk is based on a recent study~\cite{Wang:2024byt} which has extracted the TFFs from the nucleon ground state to twelve $N^*$ and $\Delta$ states. The TFFs are defined at the resonance poles, in the framework of a comprehensive DCC approach -- the Juelich-Bonn-Washington model~\cite{Mai:2021vsw,Mai:2021aui,Mai:2023cbp}. In this model the hadronic final-state interactions are described with interaction potentials iterated in Lippmann-Schwinger-like equations (for details, see e.g. Ref.~\cite{Wang:2022osj}), while the electro- and photoproduction amplitudes are constructed from the hadronic final-state interactions by Watson's final state theorem, with extra constraints such as Siegert's theorem. In a previous study~\cite{Mai:2023cbp}, the multipole amplitudes of $\pi N$, $\eta N$, and $K\Lambda$ coupled-channel electroproductions have been determined from the fits to $10^5$ data points, with about $5\times 10^4$ extra data points at the physical photon point as boundary conditions~\cite{Ronchen:2022hqk}. The center-of-mass energy ranges from $1.13$ GeV to $1.8$ GeV, and the virtuality $Q^2\in[0,8]$ GeV$^2$. Owing to the existing ambiguities in the experimental data~\cite{Smith:2023gku}, efforts have been made in Ref.~\cite{Mai:2023cbp} to fully to explore the parameter space, resulting in four fit solutions at different local minima of the $\chi^2$ that reflects the uncertainties from the data set. That provides also the estimates of the errors in the extracted TFFs. 

Technically, the TFFs are related to the residues from the following Laurent expansion at the resonance pole ($z_p$): 
\begin{equation}
  \mathcal{H}_{h}^{l\pm,I}=\frac{\widetilde{\mathcal{H}}_{h}^{l\pm,I}}{z-z_p}+\cdots\ ,\quad
	\tau^{l\pm,I}=\frac{\widetilde{R}^{l\pm,I}}{z-z_p}+\cdots\,,
\end{equation}
where $\mathcal{H}$ denotes the multipole amplitude (either $\mathcal{A}$ or $\mathcal{S}$), $\tau$ is the $\pi N$ scattering amplitude, and $\widetilde{\mathcal{H}}$, $\widetilde{R}$ are residues. Following Ref.~\cite{Workman:2013rca}, the TFFs $H_h^{l\pm,I}$, i.e. $A_{1/2},\,A_{3/2},\,S_{1/2}$, which are transition amplitudes in helicity basis, are defined in a reaction independent way via the residues: 
\begin{equation}
\label{TFFdef}
	H_{h}^{l\pm,I}(Q^2)=
	C_I\sqrt{\frac{p_{\pi N}}{\omega_0}\frac{2\pi(2J+1)z_p}{m_N\widetilde{R}^{l\pm,I}}}\widetilde{\mathcal{H}}_{h}^{l\pm,I}(Q^2)\,,
\end{equation}
with $\omega_0$ the energy of the photon at $Q^2=0$, $m_N$ the nucleon mass, $p_{\pi N}$ the three momentum of $\pi N$, and the isospin factor $C_{1/2}=-\sqrt{3}$ and $C_{3/2}=\sqrt{2/3}$. 

\begin{figure}[t]
\centering
\includegraphics[width=0.31\linewidth,trim=0.5cm 0 0.5cm 0]{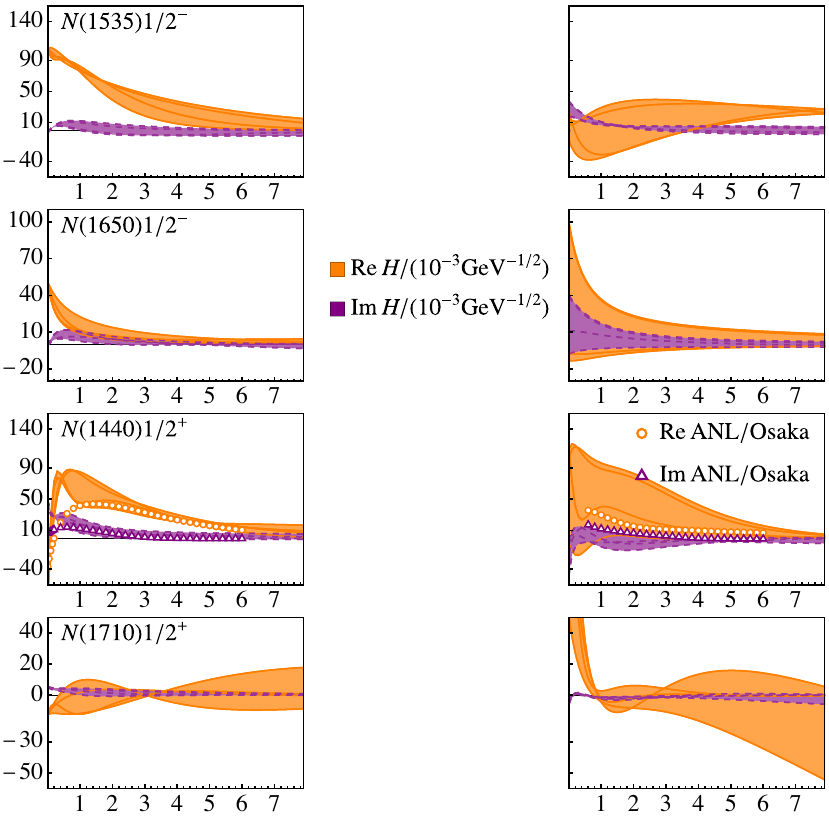}
\hspace{0.2cm}
\includegraphics[width=0.31\linewidth,trim=0.5cm 0 0.5cm 0]{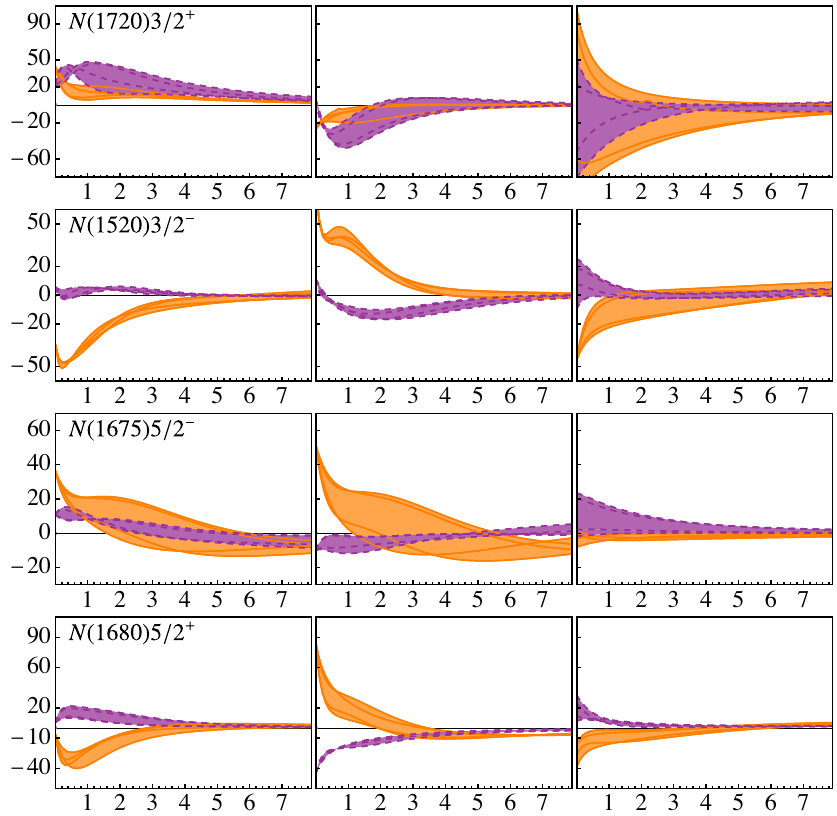}
\hspace{0.2cm}
\includegraphics[width=0.31\linewidth,trim=0.5cm 0 0.5cm 0]{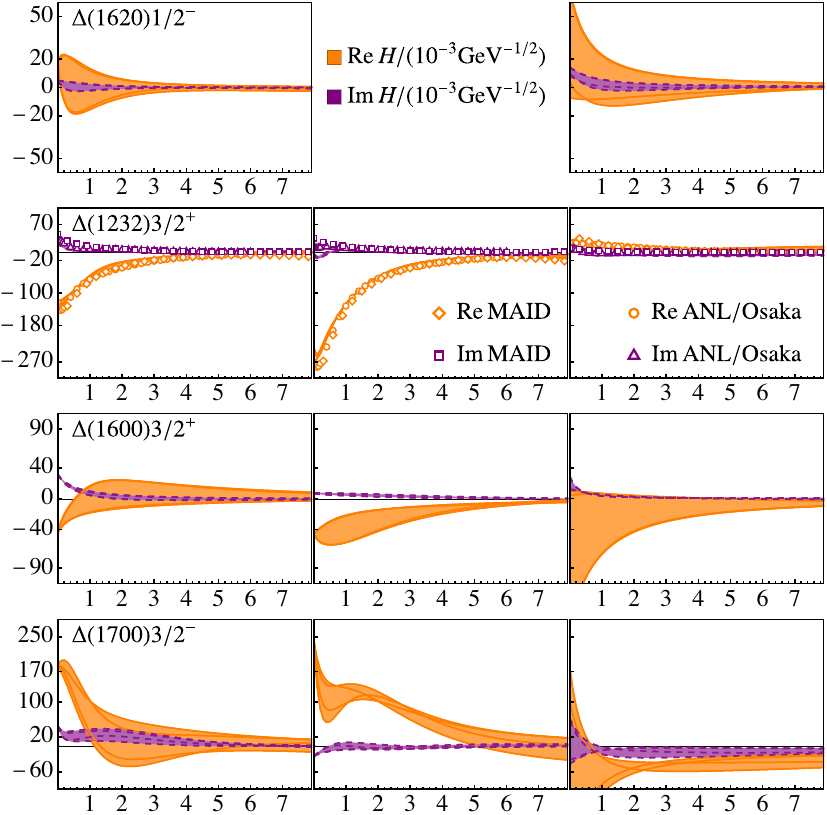}
\caption{The TFFs of the $N^*$ (left and middle panels) and $\Delta$ (right panel) states. The horizontal axis labels $Q^2/{\rm GeV}^2$. In each panel, the three figures from the left to the right shows $A_{1/2}$, $A_{3/2}$, and $S_{1/2}$, respectively. The bands represent the uncertainties stemming from the analyses in Ref.~\cite{Mai:2023cbp}. Literature results from MAID~\cite{Tiator:2016btt} and ANL-Osaka~\cite{Kamano:2018sfb} are depicted by empty symbols. }
\label{fig:TFFs} 
\end{figure}

The extracted TFFs for twelve states, namely $N(1535)$, $N(1650)$, $N(1440)$, $N(1710)$, $N(1720)$, $N(1520)$, $N(1675)$, $N(1680)$, $\Delta(1620)$, $\Delta(1232)$, $\Delta(1600)$, and $\Delta(1700)$, are plotted in Fig.~\ref{fig:TFFs}. The TFFs of $N(1440)$ and $\Delta(1232)$ are frequently discussed in the literature. For $\Delta(1232)$, the uncertainties are rather small. For comparison, the results from the ANL-Osaka mode~\cite{Kamano:2018sfb} and the MAID model~\cite{Tiator:2016btt} are also shown. The real parts are in good agreement with the result here within the uncertainties. Some discrepancies can be seen in the curves of $\text{Im}\,A_{3/2}$ in the small-$Q^2$ region, which are due to the difference at the photon-point ($Q^2=0$). 

\begin{figure}[t]
\centering
\includegraphics[width=0.35\linewidth,trim=0.5cm 0 0.5cm 0]{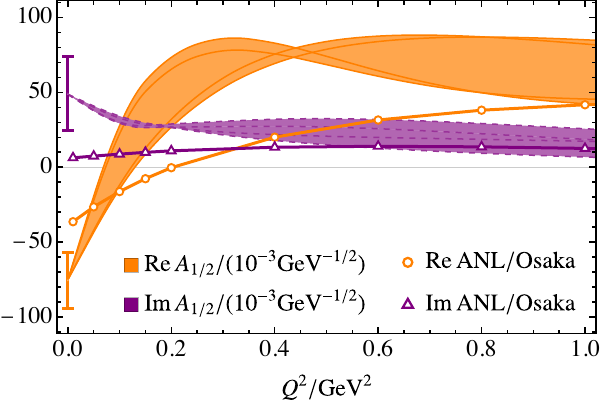}
\hspace{1cm}
\includegraphics[width=0.35\linewidth,trim=0.5cm 0 0.5cm 0]{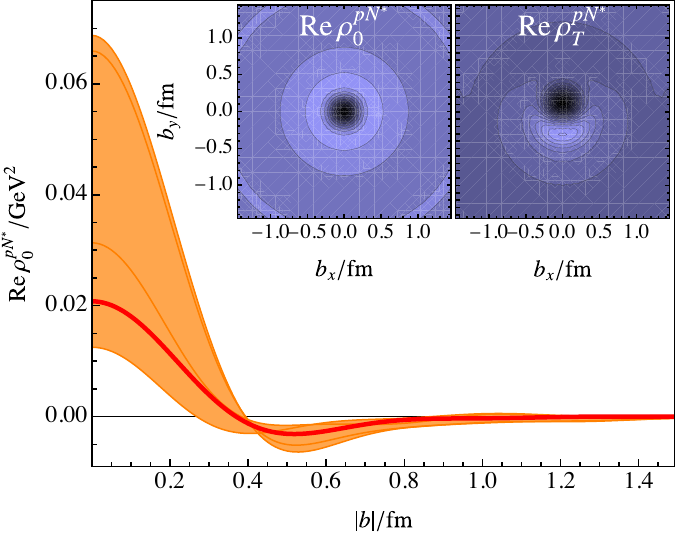}
\caption{The amplified plot of the TFFs of $N(1440)$ in the small-$Q^2$ region (left), and the transverse transition charge distributions of $N(1440)$ (right). For the figure on the right: the inset shows corresponding coordinate decompositions with light/dark shades representing negative/positive values; the orange band (thick red line) depicts the result of $\rho_0^{pN^*}$ shown in this talk (from the MAID 2007 solution~\cite{Drechsel:2007if}). }
\label{fig:1440} 
\end{figure}

For the $N(1440)$ state, see Fig.~\ref{fig:1440}, the result here reproduces a zero crossing in the ${\rm Re}\,A_{1/2}$ TFF, which implies that the core of $N(1440)$ can be explained as a radial excitation of the nucleon~\cite{Ramalho:2023hqd}. However, the structure of the $N(1440)$ is complicated: the contributions from the meson clouds or two-hadron systems seem not to be negligible~\cite{Meissner:1984un,Wang:2023snv}. According to Ref.~\cite{Tiator:2009mt}, the TFF of $N(1440)$ can also be transformed into the transverse transition charge distributions: unpolarized $\rho_0^{pN^*}$, and polarized along x-axis $\rho_T^{pN^*}$, which are also shown in Fig.~\ref{fig:1440}. The result here is in qualitative agreement with the MAID 2007 solution~\cite{Drechsel:2007if}. 

For the other states such as $\Delta(1600)$, there are no previous results also defined at the resonance poles. Direct comparisons are not made here to some latest literature results from the Breit-Wigner parametrization, e.g. in Ref.~\cite{Mokeev:2023zhq}. However, huge disagreements are not expected even if the real parts here are naively compared to them. 

To summarize, this talk presents the electromagnetic transition form factors from the nucleon to twelve $N^*$ and $\Delta$ states in Ref.~\cite{Wang:2024byt}, which are extracted from more than $10^5$ electroproduction data points by the Juelich-Bonn-Washington model. The $\pi N$, $\eta N$, and $K\Lambda$ coupled channels have been considered, with the form factors defined from the residues at the resonance poles. The results of $\Delta(1232)$ and $N(1440)$ shown here are in qualitative agreement with previous literature results, while there are no literature results (defined at the pole) for the other states. In the future the center-of-mass energy in this model will be extended to $1.9$ GeV in order to extract the transition form factors of more high-lying resonances. 

{\it Acknowledgements. }The speaker would like to thank the collaborators in the Juelich-Bonn-Washington Collaboration: Michael D{\"o}ring, Jackson Hergenrather, Maxim Mai, Terry Mart, Ulf-G.~Mei{\ss}ner, Deborah R{\"o}nchen, and Ronald Workman. The granted computing time on the supercomputer JURECA~\cite{JURECA} at Forschungszentrum J\"ulich under grant no. ``baryonspectro" is also acknowledged. This talk is supported by the National Natural Science Foundation of China under Grant No. 12175240 and the Fundamental Research Funds for the Central Universities. 
\bibliographystyle{JHEP}
\bibliography{hadron2025.bib}

\end{document}